\documentstyle[12pt,aaspp4]{article}
\begin{document}
\newcommand{\half}{\case{1}{2}} 
\newcommand{\third}{\case{1}{3}} 
\newcommand{\NN}{{\cal N}} 
\newcommand{\Nz}{{\cal N}_0} 
\newcommand{\No}{{\cal N}_1}
\newcommand{\Np}{{\cal N}_+}
\newcommand{\Nm}{{\cal N}_-}
\newcommand{\be}{\begin{equation}} 
\newcommand{\ee}{\end{equation}}
\newcommand{\en}{\varepsilon} 
\newcommand{\KK}{{\cal K}} 
\newcommand{\Fl}{{\cal F}} 
\title{An Analytic Approximation for Plane Parallel Compton Scattering Near Accretion Disks} 
\author{Robert V. Wagoner\altaffilmark{1}} 
\affil{Department of Physics and Center for Space Science and Astrophysics \\ 
Stanford University, Stanford, CA 94305--4060 \\ and Institute for Theoretical Physics, U.C. Santa Barbara}
\and\author{Alexander S. Silbergleit\altaffilmark{2}}
\affil{Gravity Probe B, Stanford University, Stanford, CA 94305--4085}
\altaffiltext{1}{wagoner@leland.stanford.edu}
\altaffiltext{2}{gleit@relgyro.stanford.edu \\[0.5cm] 
Preprint NSF--ITP--99--041. Accepted for publication by The Astrophysical Journal on July 8, 1999.}

\begin{abstract}

Compton scattering by a locally plane parallel atmosphere is analyzed within the Kompaneets approximation. In contrast to almost all previous analytic calculations of inverse Compton scattering, we assume neither homogeneity nor isotropy of the photon distribution. However, we do adopt an assumed angular distribution of the photons. For any specified energy dependence of the flux entering the base of the atmosphere, an analytic expression is obtained for the flux at any optical depth.  Only integrals over the input flux are involved. If the thickness of the atmosphere is much less than its radial position on an accretion disk, the contribution from all radii may be summed. The initial motivation (addressed in a subsequent paper) is the computation of the energy spectrum of (possibly modulated) photons originating from a specified radial region of an accretion disk, relative to the total energy spectrum. 

\end{abstract}


\newpage
\section{General Formulation}

Let us consider Compton scattering of photons by an atmosphere of thermalized hot electrons above a plane surface (such as an accretion disk) from which the photons were injected. Unlike most previous calculations, we shall not assume that the photon distribution is either isotropic in direction or homogeneous in space. 

We shall make a local approximation that gradients normal to the surface (in the $z$ direction) are dominant. However, we shall allow the properties of the atmosphere (such as electron temperature $T_e$) to vary more slowly in the radial direction along the surface.
Eventually, the contributions from a range of radii will be summed.  A useful presentation of the essentials of Compton scattering is provided by Rybicki \& Lightman \markcite{RL}(1979). A semi-analytical investigation of plane-parallel inverse Compton scattering has been carried out by Haardt \markcite{H}(1993). 

We also assume no explicit time dependence, and gravity weak enough to conserve photon momentum between collisions. The Boltzmann equation governing the photon occupation number $\NN(z,\theta,E)$ (phase space density $\times\; h^3$) is then 
\be 
\frac{D\NN}{Dt} = \vec{v}\cdot\nabla\NN = c\frac{\partial\NN}{\partial z}\cos\theta =
\left(\frac{D\NN}{Dt}\right)_{coll.}  \; , \label{trans} 
\ee 
where $\theta$ is the angle between the photon direction and the $z$ axis and $E$ is its energy. Note that the spatial gradient replaces the time derivative that is usually employed. We shall take the angular distribution of the photons to be given by 
\be 
\NN = \Nz(z,E) + \No(z,E)\cos\theta \; , \label{ang} 
\ee 
appropriate for a nearly isotropic distribution (as in the diffusion approximation).  The photon energy flux (per unit energy, in the $z$ direction) is then 
\be 
\Fl_E = (8\pi/3)c^{-2}h^{-3}E^3\No \; .  \label{flux} 
\ee 
We shall see that our governing equations will also be valid for other assumed angular distributions, in particular the two-stream approximation.

In addition we assume that the electrons have no bulk motion and are nonrelativistic ($kT_e\ll m_ec^2$).  We also invoke the Kompaneets (\markcite{KO}1957) approximation that the change in the photon energy in a scattering event is much less than $kT_e$.  This is equivalent to the condition 
\be 
\frac{kT_e}{m_ec^2}|(4-\en)\en| \ll 1 \; , \quad\quad \en\equiv\frac{E}{kT_e(z)} \; , \label{defKomp} 
\ee 
which also implies that $E\ll m_ec^2$. We note that in this case of nonrelativistic thermal electrons, the analysis of Haardt (\markcite{H}1993) invoked the additional assumption $E\ll kT_e$ (which we do not invoke in our general analysis below). 

Neglecting other processes (such as bremsstrahlung), the collision term then becomes
\be 
\left(\frac{D\NN}{cDt}\right)_{coll.}  = \chi(z)\{[-1 + {\cal O}(kT_e/m_ec^2) + {\cal O}(E/m_ec^2)]\No\cos\theta +  \case{1}{3}\KK(\Nz-\case{2}{5}\No\cos\theta)\} \; .  \label{coll} 
\ee 
Here $\chi(z) = n_e(z)\sigma_T$, the Thompson electron scattering opacity. The Kompaneets contribution to the collision term has been expressed in terms of the reduced Kompaneets operator $\KK$, given by:  
\be 
\KK(\NN) \equiv 3\frac{kT_e(z)}{m_ec^2}\frac{1}{\en^2}\frac{\partial}{\partial\en} 
\left[\en^4\left(\frac{\partial\NN}{\partial\en}+\NN\right)\right] \; .  \label{Komp} 
\ee 
In this expression, we have neglected the stimulated scattering contribution ${\NN\,}^2$, equivalent to the assumption $\NN\ll 1$. Since the operator $\KK\sim\max(kT_e/m_ec^2,E/m_ec^2)$, we see that we can also neglect it in the portion of the collision term (\ref{coll}) that is proportional to $\cos\theta$.

We introduce the optical depth $d\tau = -\chi(z)\, dz$.  The zeroth and first angular moments of the transfer equation (\ref{trans}) can then be combined to give our basic transfer equations 
\begin{eqnarray}
\partial\Nz/\partial\tau & = & \No \; , \label{first} \\ 
\partial^2\Nz/\partial\tau^2 & = & -\KK(\Nz) \; .  \label{second} 
\end{eqnarray} 
Of course, it follows that the total number flux of photons is conserved: $\int_0^\infty\No E^2dE$ is independent of optical depth $\tau$. The diffusion term on the left hand side of equation (\ref{second}) was introduced by Katz (\markcite{KA}1976), who obtained numerical solutions for both a central and a uniform photon source within a uniform spherical cloud of hot electrons. However, he assumed that the radiation field was isotropic. 

We adopt the following boundary conditions.  At the base of the atmosphere ($\tau=\tau_*$) we specify the incoming flux (or equivalently, $\No$).  At the top of the atmosphere ($\tau=0$) we specify the usual stellar atmosphere relation \markcite{MI}(Mihalas 1978) between flux and energy density (or equivalently, $\No$ and $\Nz$) that follows from the assumption of no incoming radiation.  These boundary conditions assume the form 
\begin{eqnarray} 
\No(\tau_*,E) & = & F(E) \; , \label{bound1} \\ 
\No(0,E) & = & C^{-1}\Nz(0,E) \; , \label{bound2} 
\end{eqnarray} 
where $F(E)$ is a specified function and $C\sim 1$ is a specified constant.  In the energy domain, we demand that the photon energy density (per unit energy) vanish at zero energy and the total photon energy be bounded. These requirements are satisfied if, in particular,
\begin{eqnarray} 
\Nz & = & {\cal O}(E^\alpha) \; , \; \alpha > -3 \quad \mbox{as } E\rightarrow 0 \; , \label{E0} \\ 
\Nz & = & {\cal O}(E^\beta) \; , \;  \beta < -4 \quad \mbox{as } E\rightarrow \infty \;  \label{Einf}
\end{eqnarray}
(the last condition includes exponential decay at infinity). Conditions (\ref{E0}) and (\ref{Einf}) also apply to $\No$.

Consider adopting the two-stream angular distribution
\be
\NN = \Np(z,E)\Theta(\cos\theta) + \Nm(z,E)\Theta(-\cos\theta) \label{ts}
\ee
in place of equation (\ref{ang}); where the step function $\Theta(x<0)=0,\ \Theta(x>0)=1$. 
Then it can be shown that the key equations (\ref{flux}), (\ref{Komp}), 
(\ref{first}), (\ref{second}) and the boundary conditions (\ref{bound1}), (\ref{bound2}), (\ref{E0}), and (\ref{Einf}) retain the same form. In fact, they can be obtained from the substitutions
\be
\Nz\rightarrow \half(\Np + \Nm) \; , \quad \No\rightarrow \case{3}{4}(\Np - \Nm) \; , \quad C\rightarrow 2/3 \; .
\ee
The two-stream approximation should be more accurate for $\tau_*\ll 1$, while the angular distribution (\ref{ang}) should be more accurate for $\tau_*\gg 1$. It is comforting to know that this formalism includes both cases.

\section{General Solution}

It is helpful to introduce a reduced phase space density $\Phi(\tau,E)$ by removing the solution corresponding to pure Thompson scattering ($\partial^2\Nz/\partial\tau^2=0$), so that
\be
\Nz \equiv (C+\tau)F(E) + \Phi(\tau,E)  \label{red}
\ee
and $\No = F(E) + \partial\Phi/\partial\tau$.
The boundary conditions (\ref{bound1}) and (\ref{bound2}) then give
\be
\frac{\partial\Phi}{\partial\tau}(\tau_*,E) = 0 \; , \quad\quad
C\frac{\partial\Phi}{\partial\tau}(0,E) - \Phi(0,E) = 0 \; , \label{bound}
\ee
while the master equation (\ref{second}) becomes
\be
\partial^2\Phi/\partial\tau^2 + \KK(\Phi) = -(C+\tau)\KK(F) \; . \label{master2}
\ee

We now assume that $T_e(z) =$ constant, in which case the homogeneous form of this equation ($F=0$) has separated 
solutions $\Phi = Q(\tau)P(\en)$, with $d^2Q/d\tau^2 + \kappa^2Q = 0$ and $\kappa^2$ the separation constant. [Recall from equation(\ref{defKomp}) that the relative energy $\en=E/kT_e$.] We expand solutions to the master equation (\ref{master2}) in terms of the corresponding eigenfunctions:
\be
\Phi = \sum_n f_n(\en)Q_n(\tau) \; , \label{expand}
\ee
where the eigenfunctions $Q_n$ and eigenvalues $\kappa_n$ that satisfy the boundary conditions (\ref{bound}) are determined by
\be
Q_n = A_n\cos[\kappa_n(\tau_*-\tau)] \; , \quad\quad \tan(\kappa_n\tau_*) = 1/(C\kappa_n) 
\quad (n=1,2,3,\ldots) \; .  \label{eigen}
\ee
Imposing the normalization $\int_0^{\tau_*}Q_m(\tau)Q_n(\tau)\, d\tau = \delta_{mn}$ gives
\be
A_n^2 = 2[\tau_* + C\sin^2(\kappa_n\tau_*)]^{-1} \; .
\ee 
We note that $\kappa_n$ must be real.
 
The master equation (\ref{master2}) then reduces to the ordinary differential equation
\be
\KK(f_n) - \kappa_n^2f_n + \kappa_n^{-2}A_n\KK(F) = 0 \; , \label{master1}
\ee
using the orthonormality of $Q_n(\tau)$.
Finally, we replace the function $f_n(\en)$ by $g_n(\en)$ according to $f_n(\en)=A_n\left[ g_n(\en)- \kappa_n^{-2}F(\en)\right]$, which allows us to simplify the source term (involving $F$) in equation (\ref{master1}). Using the definition (\ref{Komp}) of the differential operation $\KK$, this now assumes the explicit form
\be
\left[\en^2\frac{d^2}{d\en^2} + (\en^2+4\en)\frac{d}{d\en} + (4\en-\lambda^2\kappa_n^2)\right]g_n(\en) = -\lambda^2F(\en) \; ,  \label{master0}
\ee
where
\be
\lambda^2 \equiv m_ec^2/3kT_e \quad (\gg 1) \; .
\label{lamb}  
\ee
Combining the above results, we obtain 
$$
\No(\tau,\en) = \frac{\partial\Nz}{\partial\tau}(\tau,\en) = F(\en)+\frac{\partial\Phi}{\partial\tau}(\tau,\en)
$$
\be
= F(\en) + \sum_nA_n^2\kappa_n\sin[\kappa_n(\tau_*-\tau)][g_n(\en) - \kappa_n^{-2}F(\en)]=
 \sum_nA_n^2\kappa_n\sin[\kappa_n(\tau_*-\tau)]g_n(\en)\; .  \label{rel}
\ee
[The last series converges nonuniformly near $\tau=\tau_*$, such that one obtains the expected result: $\No(0,\en)\rightarrow F(\en)$ as $\tau_*\rightarrow 0$.] This equation relates the major quantity of observable interest [the flux $\Fl_E(0,E;r)\propto E^3 \No(0,\en)$] to the solution $g_n(\en)$ to our reduced master equation (\ref{master0}), for any input flux $F(\en;r)$. In this context, the radius $r$ is a parameter that we later integrate over to obtain the corresponding luminosity. 

Now consider the homogeneous part of equation (\ref{master0}). Its independent solutions are $h_1(\en) = C_1e^{-\en}\en^s M(s,2s+4,\en)$ and $h_2(\en) = C_2e^{-\en}\en^s U(s,2s+4,\en)$, where $M(a,b,x)$ and $U(a,b,x)$ are confluent hypergeometric (Kummer) functions \markcite{AS}(Abramowitz \& Stegun 1972). The index $s$ assumes the values
\be
s_\pm(n) = \frac{3}{2}\left(-1\pm\sqrt{1+\frac{4}{9}\lambda^2\kappa_n^2}\right) \; .\label{spm}
\ee
We note that as $\en\rightarrow 0$, the function $M(s,2s+4,\en)\rightarrow 1$ while
\begin{eqnarray}
U(s,2s+4,\en) & \rightarrow &\frac{\Gamma(2s+3)}{\Gamma(s)}\en^{-(2s+3)} \; \mbox{ for } s=s_+(n)(>0) \\  & \rightarrow &\frac{\Gamma(-2s-3)}{\Gamma(-s-3)} \quad\quad\;\;\:\mbox{ for } s=s_-(n)(<-3) \; . \label{U0}
\end{eqnarray}
As $\en\rightarrow\infty$, these functions obey
\be      
M(s,2s+4,\en)\rightarrow\frac{\Gamma(2s+4)}{\Gamma(s)}e^\en\en^{-s-4} \; , \quad
U(s,2s+4,\en)\rightarrow\en^{-s} \; . \label{MU}
\ee

The general solution to equation (\ref{master0}) is $g(\en) = u_1(\en)h_1(\en) +  u_2(\en)h_2(\en)$, with $u_a(\en)$ found by standard techniques. The boundary conditions (\ref{E0}) and (\ref{Einf}) plus the above asymptotic behaviors of the confluent hypergeometric functions require that $u_1\rightarrow 0$ as $\en\rightarrow\infty$ and $u_2\rightarrow 0$ as $\en\rightarrow 0$, while also demanding that we choose $s=s_n\equiv s_+(n)$. 
The solution to our master equation (\ref{master0}) is then completely specified as
\be
g_n(\en) = \frac{\lambda^2\Gamma(s_n)e^{-\en}\en^{s_n}}{\Gamma(2s_n+4)} \left[M_n(\en)\int_\en^\infty U_n(\en_*)F(\en_*)\en_*^{s_n+2}d\en_* 
+ U_n(\en)\int_0^\en M_n(\en_*)F(\en_*)\en_*^{s_n+2}d\en_*\right] \; , \label{sol}
\ee
where we have adopted the final convenient notations
\be
M_n(\en)\equiv M(s_n,2s_n+4,\en) \: , \qquad U_n(\en)\equiv U(s_n,2s_n+4,\en) \; .\label{hyper}
\ee

\section{Application to Thin Accretion Disk Atmospheres}

We now apply the developed formalism to a hot atmosphere above the surface of an accretion disk which feeds a compact object (black hole, neutron star, or white dwarf). The key assumption made is that the scale of radial variation of the properties of the disk and atmosphere is much larger than the effective physical thickness of the atmosphere. We can then apply the above results separately to each interval $dr$ of radial extent. The energy flux $\Fl_E(E)$ from the surface of the disk into the base of the atmosphere (at optical depth $\tau_*$) is related to the source function $F(E)$ through equations (\ref{flux}) and (\ref{bound1}).

The standard semirelativistic thin disk model \markcite{SS}(Shakura \& Sunyaev 1973) is adopted. The input flux is approximated as a (usually dilute) black body characterized by a color temperature $T_d(r)$, which corresponds to 
\be
F(E,r) = F_0[e^{E/kT_d(r)} - 1]^{-1} \; , \label{input}
\ee
where the constant $F_0\le 1$ could also be a slowly varying function of $r$. [However, our assumption $\Nz\ll 1$ requires the same of $F(E,r)$.] 

Let us adopt the modified Newtonian potential $\Phi = -(GM/r)[1-3(GM/rc^2)+12(GM/rc^2)^2]$ \markcite{NW}(Nowak \& Wagoner 1991). The disk temperature is then \markcite{N}(Nowak 1992)
\be
T_d(r) = 10.286 T_{max}r^{-3/4}\left(1-\frac{8}{r}+\frac{60}{r^2}\right)^{1/4}
\left(1-\sqrt{\frac{6}{r-6+36/r}}\right)^{1/4} \; .
\ee
The (dimensionless) radial coordinate $r$ is now in units of $GM/c^2$, with $M$ the mass of the (slowly rotating) central compact object. The inner radius of the disk ($r_i$) will be at least that of the innermost stable circular orbit, $r=6$. The maximum temperature $T_{max}$ of the disk is reached at the radius $r=9.795$. 

We also allow the electron temperaure and the total optical depth of the atmosphere to vary slowly with $r$, and be given by the functions
\be
T_e(r) = T_0(r/r_i)^{-a} \; , \quad\quad \tau_*(r) = \tau_0(r/r_i)^{-b} \; . \label{func}
\ee
We will also specify an outer radius $r_o$ of the atmosphere. Since often we shall also be  interested in the emergent energy spectrum of (possibly modulated) photons originally injected from a small radial region of the disk, we must specify its inner and outer radii $r_1$ and $r_2$ in that case.

In summary, a given model corresponds to specification of the parameters \{$C,\ a,\ b,\ r_i,\ r_o,\ r_1,\ r_2,\ F_0,\ T_{max},\ T_0,\ \tau_0$\}. The output of interest is the total and modulated luminosity per unit energy (from both sides of the disk):
\be
L_E = 4\pi\int_{r_i}^\infty\Fl_E(0,E;r)\, r\, dr \; , \quad \tilde{L}_E = 4\pi\int_{r_1}^{r_2}\Fl_E(0,E;r)\, r\, dr  \; . \label{lum}
\ee

\section{Effective Solution}

The key to the effective solution lies in the simplification of the expression (\ref{sol}) for $g_n(\en)$ in various ranges of energy. We shall carry out such simplification for the input flux (\ref{input}) from a dilute black body,
\be
F(\en)=F_0\,\left(e^{\en/\alpha}-1\right)^{-1} \; , \qquad \alpha(r)\equiv T_d(r)/T_e(r) \; .
\label{inpep}
\ee
First we consider the extreme low energy limit $E\ll kT_d\, ,\, E\ll kT_e$. Using the appropriate asymptotics of the confluent hypergeometric functions (\ref{hyper}) and expression (\ref{inpep}), it is straightforward to show that to lowest order,
\be
g_n(\en)=\frac{\lambda^2}{(s_n+1)(s_n+2)}\,\frac{\alpha}{\en}\; , \qquad \en\ll\min(1,\alpha) \; , \label{low}
\ee
with the expression for $s_n=s_+(n)$ given in equation (\ref{spm}). Both integrals in equation (\ref{sol}) contribute to this result.

The extreme high energy limit $E\gg kT_e$ is treated in a similar fashion under the additional condition $T_d < T_e$, except that the main contribution comes here solely from the second integral in equation (\ref{sol}):
\be
g_n(\en)=\lambda^2 F_0\,\frac{\Gamma(s_n)K_n(\alpha)}{\Gamma(2s_n+4)}\, e^{-\en} \; , \qquad
\en\gg 1 \; , \; \alpha < 1 \; , \label{high}
\ee
with
\be
K_n(\alpha)=\int_0^\infty\frac{M_n(\en)\en^{s_n+2}d\en}{e^{\en/\alpha}-1} \; . \label{const}
\ee
The integral converges at infinity because $\alpha<1$.

These results still leave us to cope with the whole range of intermediate energies. The gap can be covered for the observationally relevant situation when 
\be
\alpha(r)= T_d(r)/T_e(r) \ll 1 \; . \label{limit}
\ee
This inequality is valid for essentially all models of accretion disk systems with hot  `coronae', and in all such cases  the following approximation holds to lowest order in $\alpha$:
\be
g_n(\en) = 8\pi^{1/2}\lambda^2F_0\,\frac{(s_n+2)\Gamma(s_n)\zeta(s_n+3)}{\Gamma(s_n+5/2)}\,
\left(\frac{\alpha}{4}\right)^{s_n+3}\en^{s_n}U_n(\en)e^{-\en}\; , \quad \alpha\ll\min(1,\en) \; . \label{interm}
\ee
This expression contains the Riemann zeta function $\zeta(z)=\sum_{k=1}^\infty k^{-z}\; (\mbox{Re}\, z > 1)$. The derivation of equation (\ref{interm}) is rather complicated and thus is given in the Appendix.

From equation (\ref{spm}) and the fact that $(n-1)\pi < \kappa_n\tau_* < (n-1/2)\pi$ from equation(\ref{eigen}), we see that 
\be
s_n\approx \lambda\kappa_n\approx n\pi\lambda/\tau_* \; , \quad n\rightarrow\infty \; .  
\ee
Note that the value of the expression (\ref{interm}) drops very rapidly with $n$ not only because of this increasing power of the small parameter $\alpha$, but also due to the factor in front of it that decreases rapidly with $n$. Note also that for $\en\gg 1$ this formula agrees with the high energy limit (\ref{high}) for $\alpha\ll 1$.

Let us now use the above expressions to obtain the flux function $\No(\en,\tau)$ in the corresponding energy ranges. From equation (\ref{rel}) we find that
\begin{eqnarray}
\No(\en,\tau) & = & \lambda^2 F_0\,\frac{\alpha}{\en}\,N_-(\tau)\; , \quad \en\ll\min(1,\alpha)\; , \label{I1} \\
N_-(\tau) & \equiv & \sum_n \frac{A_n^2\kappa_n\sin[\kappa_n(\tau_*-\tau)]}{(s_n+1)(s_n+2)}\; ;  \nonumber \\
\No(\en,\tau) & = & \lambda^2 F_0N_0\left(\frac{\alpha}{4}\right)^{s_1+3}\en^{s_1}U_1(\en)e^{-\en}\sin[\kappa_1(\tau_*-\tau)]\; , \quad \alpha\ll\min(\en,\en^{-1})\; , \label{I2} \\
N_0 & \equiv & 8\pi^{1/2}\,\frac{\left(s_1+2\right)\Gamma(s_1)\zeta(s_1+3)A_1^2\kappa_1}{\Gamma(s_1+5/2)}\; ;
\nonumber \\
\No(\en,\tau) & = & \lambda^2 F_0e^{-\en}N_+(\tau)\; , \quad \en\gg 1\; , \quad \alpha<1\; , \label{I3} \\ 
N_+(\tau) & \equiv & \sum_n\frac{\Gamma(s_n)K_n(\alpha)A_n^2\kappa_n\sin[\kappa_n(\tau_*-\tau)]}{\Gamma(2s_n+4)} \; . \nonumber  
\end{eqnarray}
In order that the first term in the series dominates, giving equation (\ref{I2}), we have had to introduce the additional restriction $\alpha\en\ll 1$. We also note that to lowest order in $\alpha\ll 1$, the integral $K_n(\alpha) = \Gamma(s_n+3)\zeta(s_n+3)\alpha^{s_n+3}$ in equations (\ref{high}) and (\ref{I3}). Thus mainly $n=1$ contributes, and we recover the high energy limit of the intermediate equations (\ref{interm}) and (\ref{I2}).

We see that for any given value of the optical depth $\tau$ the energy distribution consists of the extreme low energy (Rayleigh--Jeans) part $\No\propto\alpha/\en$ and the extreme high frequency (Wien) part $\No\propto\exp(-\en)$. For values of the temperature ratio $\alpha\ll 1$, the whole intermediate range of energies is described by the dependence
$$
\No\propto\en^{s_1}U_1(\en)e^{-\en},
$$
which turns to $\No\propto \en^{-(s_1+3)}$ for $\alpha\ll\en\ll 1$ and to the decaying exponential (\ref{I3}) for $\en\gg 1$. Recall from equation (\ref{spm}) that
\be
s_n = \frac{3}{2}\left(-1+\sqrt{1+\frac{4}{9}\lambda^2\kappa_n^2}\right) \label{snp}
\ee
is specified by the electron rest mass to thermal energy ratio $\lambda^2\gg 1$ [equation (\ref{lamb})], the total optical depth $\tau_*$ and the boundary condition parameter $C$ [the latter two through the eigenvalue $\kappa_n^2$ satisfying equation (\ref{eigen})].

In order to understand the behavior of the major observables, the luminosities given by equation (\ref{lum}), let us exhibit the dependence of the essential input, the emergent photon energy flux $\Fl_E(0,E;r)$, on the relevant physical quantities $E$, $F_0(r)$, $T_e(r)$, $T_d(r)$, and $\tau_*(r)$. The last three induce the radial dependences of $\en(r)$, $\alpha(r)$, $\lambda(r)$, $\kappa_n(r)$, $s_n(r)$, and $A_n(r)$.

Using equation (\ref{flux}), we define a reduced flux $\tilde{\Fl}$ from
\be
\Fl_E(0,E;r) = (8\pi m_e c^2/9c^2 h^3)F_0(r)\tilde{\Fl}(E;r) \; . \label{redF}
\ee
Thus $\tilde{\Fl}$ has units of (energy)${}^2$. From equations (\ref{I1}), (\ref{I2}), and (\ref{I3}) we then obtain the dependences
\begin{eqnarray}
\tilde{\Fl}(E;r) & = & E^2[T_d(r)/T_e(r)]N_-(\tau=0; r)\; ,\quad E\ll\min[kT_d,kT_e] \; ;\label{F1} \\
\tilde{\Fl}(E;r) & = & E^{s_1(r)+3}U_1(E/kT_e(r))e^{-E/kT_e(r)}[kT_d(r)]^{s_1(r)+3}
[kT_e(r)]^{-[2s_1(r)+4]}\tilde{N}_0(r) \; , \nonumber \\
\tilde{N}_0(r)   & = & 4^{-[s_1(r)+3]}\sin[\kappa_1(r)\tau_*(r)]N_0(r) \; , \quad kT_d\ll\min[E,(kT_e)^2/E] \; ; \label{F2} \\
\tilde{\Fl}(E;r) & = & [E^3/kT_e(r)]e^{-E/kT_e(r)}N_+(\tau=0; r) \; , \quad E\gg kT_e, \; T_d < T_e
\; . \label{F3} 
\end{eqnarray}
We see that in general, the integration over the $r$ dependences of these fluxes to obtain the luminosities $L_E(E)$ and $\tilde{L}_E(E)$ [via equation (\ref{lum})] must be carried out numerically. Only for the lowest energies [equation (\ref{F1})] is the energy dependence universal, factoring out of the integral.

Let us note in particular perhaps the most observationally relevant results 
\begin{eqnarray}
\Fl_E(0,E;r) & \propto & E^{-s_1(r)} \; , \qquad\quad kT_d \ll E \ll kT_e \; ; \label{power} \\ 
             & \propto & E^3 E^{-E/kT_e(r)} \; , \; kT_d < kT_e \ll E \; ; \label{exp}
\end{eqnarray}
obtained from the corresponding dependences of $\No$ indicated above. As mentioned previously, the usual treatments of pure Compton scattering (\markcite{RL}Rybicki \& Lightman 1979) invoke no spatial gradients or anisotropies. A spatially uniform source of photons is usually employed. In place of explicit photon diffusion, a mean number of scatterings $N_s$ and the corresponding Compton $y$ parameter
\be
N_s = \max (\tau_s, \tau_s^2) \; , \qquad y = \frac{4kT_e}{m_ec^2}N_s \label{Ns}
\ee
are also employed. Here $\tau_s$ is the optical depth through the region considered (usually a sphere). The same exponential behavior as given by equation (\ref{exp}) in the higher energy range is obtained in the standard approach. 

The same power--law behavior as given by equation (\ref{power}) in the lower energy range is also obtained in the standard approach. Equating the exponent 
\[ \frac{3}{2}\left(1-\sqrt{1+\frac{16}{9y}}\right) \]
obtained in the standard analysis to the expression for $-s_1$ given by equation (\ref{snp}), we obtain the correspondence $\kappa_1^2 = 3/N_s$. 
This correspondence between the lowest eigenvalue of our problem and the mean number of scatterings is also obtained when we consider our eigenvalue equation (\ref{eigen}), which has the solutions
\begin{eqnarray}
\kappa_1^2 & \cong & 1/C\tau_*      \; , \qquad \tau_* \ll 1 \; ; \label{less} \\
           & \cong & (\pi/2\tau_*)^2 \; , \quad \tau_* \gg 1 \; . \label{more} 
\end{eqnarray}  
Note that if $\tau_s = 3C\tau_*$, the correspondence is exact for $\tau_* \ll 1$; and if $\tau_s = (2\sqrt{3}/\pi)\tau_*$, the correspondence is exact for $\tau_* \gg 1$. 

\section{Discussion}

The approach that we have taken to this problem of Compton scattering in a geometrically thin atmosphere is similar to that employed in approximate analyses of radiative transfer in ordinary plane parallel stellar atmospheres \markcite{MI}(Mihalas 1978). That is, we look for time independent solutions in a situation where the radiation field depends upon one dimension and the angle from the corresponding direction. The key ingredient was the realization that one can employ the usual Kompaneets source function (corresponding to an isotropic photon distribution) even if the radiation field is anisotropic, provided that the photon energy $E\ll m_ec^2$. By contrast (as indicated above), almost all previous analytic studies of Compton scattering (e.g., \markcite{ST}Sunyaev \& Titarchuk 1980; \markcite{T}Titarchuk 1994) have taken the photon distribution to be essentially isotropic as well as homogeneous.

In addition to the usual Kompaneets approximation, there are two other major approximations that we have made. The first is an assumed form [equation (\ref{ang}) or (\ref{ts})] for the angular dependence of the radiation field, although we found that the resulting equations for the angular moments have the same form for these two angular dependences. The second was the imposition of the surface boundary condition [equation (\ref{bound2})] via an assumed relation between the energy density and flux there, similar to the approximation often made in stellar atmosphere theory. 

In the future, we hope to extend this approach to other geometries involving injection of photons into more extended distributions of hot electrons. We also plan to employ more detailed relativistic models of (black hole or neutron star) accretion disks. In addition, we will apply this formalism to observations such as the steep energy dependence of the modulated luminosity of high frequency quasi-periodic oscillations in black-hole candidate X-ray sources (\markcite{MRG}Morgan, Remillard \& Greiner 1997). Comparing with the energy dependence of our Compton scattering models can constrain the location on the accretion disk of the source of the modulation.

\acknowledgments

This research was supported in part by NASA grants NAG 5-3102 and NAS 8-39225, and NSF grant PHY94-07194. A portion of this research was completed while R.V.W. was a member of the black hole astrophysics program at the Institute for Theoretical Physics, U.C. Santa Barbara. We thank Eric Agol, Dana Lehr, and J\"{o}rn Wilms for useful discussions, and Omer Blaes for improving the presentation.

\appendix
\section*{Appendix}

To derive the approximation (\ref{interm}), we write for convenience $\beta\equiv\alpha^{-1}\gg 1$, and suppress the index $n$ of $s_n$. With these conventions, the first integral in equation (\ref{sol}), without the factor $F_0$, becomes
$$
I_+(\en)\equiv\int_\en^\infty U_n(\en_*)\left(e^{-\beta\en_*}-1\right)^{-1}\en_*^{s+2}\,d\en_*
$$
$$
= \frac{1}{\Gamma(s)}\int_0^\infty t^{s-1}(1+t)^{s+3} \int_\en^\infty e^{-\en_*t} \left(e^{\beta\en_*}-1\right)^{-1}\en_*^{s+2}\,d\en_*dt \; . \eqno(A.1)
$$
Here we have used the standard integral representation \markcite{AS}(Abramowitz \& Stegun 1972) for the confluent hypergeometric function $U_n(\en)$, and changed the order of integration. We evaluate the inner integral by expanding the denominator in the geometric progression of $\exp(-\beta\en_*)$:
$$
\int_\en^\infty \,\frac{e^{-\en_*t}\en_*^{s+2}\,d\en_*}{e^{\beta\en_*}-1}=
\sum_{k=0}^{\infty}\int_\en^\infty\, e^{-\left[\beta(k+1)+t\right]\en_*}\en_*^{s+2}\,d\en_*
$$
$$
= \en^{s+2}\sum_{k=0}^{\infty}\frac {\exp\{-[t+\beta(k+1)]\en\}}{t+\beta(k+1)}\left\{1+ {\cal O}([\beta(k+1)]^{-1})\right\} \; .
$$
Introducing this into equation (A.1), we find
$$
I_+(\en)=\en^{s+2}\sum_{k=0}^{\infty}\frac{e^{-\beta(k+1)\en}}{\Gamma(s)}
\int_0^\infty \frac{e^{-\en t}t^{s-1}(1+t)^{s+3}\,dt}{\beta(k+1)+t}\left[1+ {\cal O}(\beta^{-1})\right]
$$
$$
= \en^{s+2}\sum_{k=0}^{\infty}\frac{e^{-\beta(k+1)\en}}{\beta(k+1)}\,
\frac{1}{\Gamma(s)}\int_0^\infty e^{-\en t}t^{s-1}(1+t)^{s+3}\,dt\left[1+ {\cal O}(\beta^{-1})\right]
$$
$$
= -\frac{1}{\beta}\en^{s+2}U_n(\en)\ln(1-e^{-\beta\en})\left[1+ {\cal O}(\beta^{-1})\right]\; ,
\eqno(A.2)
$$
where the same integral representation has been used again, and the power series of $\exp(-\beta\en)$ is summed up easily to give the logarithm in the answer. The second of the expressions in equation (A.2) is obtained from the first one by means of the following integration by parts (if $\en\gg\alpha$):
$$
\int_0^\infty \frac{e^{-\en t}t^{s-1}(1+t)^{s+3}\,dt}{\beta(k+1)+t}
$$
$$
= -\frac{\int_t^\infty e^{-\en x}x^{s-1}(1+x)^{s+3}\,dx}{\beta(k+1)+t}\Biggl|_{t=0}^{t=\infty}-
\int_0^\infty \frac{\int_t^\infty e^{-\en x}x^{s-1}(1+x)^{s+3}\,dx}{[\beta(k+1)+t]^{2}}\,dt
$$
$$
= \frac{1}{\beta(k+1)}\int_0^\infty e^{-\en x}x^{s-1}(1+x)^{s+3}\,dx\left\{1+ {\cal O}([\beta(k+1)]^{-1})\right\} \; . \eqno(A.3)
$$

The second of the integrals in equation (\ref{sol}) is transformed similarly with some moderate complications. Namely, using the integral representation for the confluent hypergeometric function $M_n(\en)$, we calculate:
$$
I_-(\en)\equiv\int_0^\en M_n(\en_*)\left(e^{-\beta\en_*}-1\right)^{-1}\en_*^{s+2}\,d\en_*
$$
$$
= \frac{\Gamma(2s+4)}{\Gamma(s)\Gamma(s+4)}\int_0^1 t^{s-1}(1-t)^{s+3}\,dt
\int_0^\en e^{\en_*t}\left(e^{\beta\en_*}-1\right)^{-1}\en_*^{s+2}\,d\en_*
$$
$$
= \frac{\Gamma(2s+4)}{\Gamma(s)\Gamma(s+4)}\int_0^1 t^{s-1}(1-t)^{s+3}\,dt
\sum_{k=0}^{\infty}\int_0^\en e^{-[\beta(k+1)-t]\en_*}\en_*^{s+2}\,d\en_* \; . \eqno(A.4)
$$
The integral over $\en_*$ is evaluated as
$$
\int_0^\en e^{-[\beta(k+1)-t]\en_*}\en_*^{s+2}\,d\en_*=
\int_0^\infty e^{-[\beta(k+1)-t]\en_*}\en_*^{s+2}\,d\en_* 
- \int_\en^\infty e^{-[\beta(k+1)-t]\en_*}\en_*^{s+2}\,d\en_*
$$
$$
= \frac{\Gamma(s+3)}{[\beta(k+1)-t]^{s+3}}-
\en^{s+2}\frac{e^{-[\beta(k+1)-t]\en}}{\beta(k+1)-t}\left\{1+ {\cal O}([\beta(k+1)]^{-1})\right\} \; .
$$
This splitting is possible because  $\en$ is bounded away from zero, $\en\geq\en_0>0$, and $\beta(k+1)\geq\beta\gg 1\geq t$. The substitution of this result into equation (A.4) allows us to obtain
$$
I_-(\en)=I_1(\en)-I_2(\en) \; ; \eqno(A.5)
$$
$$
I_1(\en)=\frac{\Gamma(2s+4)\Gamma(s+3)}{\Gamma(s)\Gamma(s+4)}\,\sum_{k=0}^{\infty}\int_0^1
\,\frac {t^{s-1}(1-t)^{s+3}\,dt}{[\beta(k+1)-t]^{s+3}} \; , \eqno(A.6)
$$
$$
I_2(\en)=\en^{s+2}\sum_{k=0}^{\infty}
e^{-\beta(k+1)\en}\frac{\Gamma(2s+4)}{\Gamma(s)\Gamma(s+4)}
\int_0^1\,\frac {e^{\en t}t^{s-1}(1-t)^{s+3}\,dt}{\beta(k+1)-t}
\left[1+{\cal O}(\beta^{-1})\right] \; . \eqno(A.7)
$$
Now we notice that the expression (A.7) for $I_2(\en)$ is similar to the first expression for $I_+(\en)$ in equation (A.2). Thus, being treated in the same way as the latter in equation (A.3) with the integration by parts, it reduces to the corresponding result 
$$
I_2(\en)=-\frac{1}{\beta}\en^{s+2}M_n(\en)\ln(1-e^{-\beta\en})\left[1+ {\cal O}(\beta^{-1})\right]\; . \eqno(A.8)
$$
This is similar to the last formula in equation (A.2), but $\en\gg\alpha$ is required here. 

Employing these results in the essential part of the solution (\ref{sol}) for $g_n(\en)$, we find that the main terms with $I_+(\en)$ and $I_2(\en)$ cancel:
$$
M_n(\en)I_+(\en) + U_n(\en)I_-(\en)=M_n(\en)I_+(\en)- U_n(\en)I_2(\en)+ U_n(\en)I_1(\en)
$$
$$
= U_n(\en)I_1(\en)[1+{\cal O}(\beta^{s+1}e^{-\beta\en})] \; , \quad 
\beta\gg 1\; , \quad \en\gg 1/\beta \; . \eqno(A.9)
$$
Finally, to find the corresponding asymptotic behavior of $I_1(\en)$ we integrate by parts once again in the same way as in equation (A.3), now in the integral (A.6):
$$
I_1(\en)=\frac{\Gamma(2s+4)\Gamma(s+3)}{\Gamma(s)\Gamma(s+4)}\sum_{k=0}^{\infty}\,
\frac {\int_0^1t^{s-1}(1-t)^{s+3}\,dt}{[\beta(k+1)]^{s+3}}\left[1+ {\cal O}(\beta^{-1})\right]
$$
$$
= \Gamma(s+3)\sum_{k=0}^\infty \, [(k+1)\beta]^{-(s+3)}\left[1+ {\cal O}(\beta^{-1})\right]
= \frac{\Gamma(s+3)\zeta(s+3)}{\beta^{s+3}}\left[1+ {\cal O}(\beta^{-1})\right]\; , \eqno(A.10)
$$
where $\zeta(z)$ is the Riemann zeta function.
Using formula (A.10) in the equality (A.9) and introducing the result into the expression (\ref{sol}) for $g_n(\en)$, we arrive at the desired approximation (\ref{interm}).

\end{document}